\documentstyle[prl,preprint,aps]{revtex}

\begin{document}
 \tolerance 50000

\draft


\title{The Density Matrix Renormalization Group, 
Quantum Groups and
Conformal Field Theory} 

\author{G. Sierra$^{1}$,  
M.A. Mart{\'\i}n-Delgado$^{2}$} 
\address{ 
$^{1}$Instituto de Matem{\'a}ticas y F{\'\i}sica Fundamental, C.S.I.C.,
Madrid, Spain. 
\\ 
$^{2}$Departamento de
F{\'\i}sica Te{\'o}rica I, Universidad Complutense. Madrid, Spain.
}

\maketitle 

\begin{abstract} 
\begin{center}
\parbox{14cm}{We present an overview of the Density Matrix
Renormalization Group and its connections to Quantum Groups, Matrix Products
and Conformal Field Theory. We emphasize some common formal structures
in all these theories. We also propose  two-dimensional extensions of the
variational matrix product ansatzs.
}

\end{center}
\end{abstract}


\section{Introduction}

The Density Matrix Renormalization Group (DMRG) is a powerful
numerical Real Space RG method introduced in 1992 by S.R. White 
which can be applied to a large variety of Quantum Lattice
Hamiltonians defined in 1d, quasi 1d (ladders) and large 
clusters \cite{W1}. 
The DMRG was originally proposed in the domain of Condensed Matter
Physics, where it has already become a standard method
specially for 1d systems, 
but its range of applicability has been extended also 
to Statistical Mechanics, polymers,
Chemical Physics, etc... (for a review 
see \cite{Dresden})  and one may expect that it will be applied
in the near future to other domains of Physics as Quantum Field
Theory, Nuclear Physics, etc.

There are by now excelent reviews on the DMRG  \cite{W2}
and other
related real space RG methods \cite{UAM}, \cite{Esco} 
so it is not
the purpose of the present contribution to duplicate material
already present in the literature.   
Instead we shall try to give an overview of the DMRG method
in order to explore  its  relationships with 
Group Theory and its quantum deformation, Conformal
Field Theory (CFT) and the Matrix Product Method. 
The relation between the DMRG and Quantum Groups was suggested
in \cite{q-group} by studying the RG method of
$q-$group invariant Hamiltonians. 
We also review the variational approach
of the DMRG in terms of the so called
matrix product (MP) ansatzs introduced
by \"{O}stlund and Rommer \cite{OR} and suggest
a 2D versions of it which may lead to a 2D
formulation  of the DMRG.

\section*{Real space RG methods: generalities}

Let us suppose we have a discrete system with $N$ sites and that
at each site there are two possible states, say
spin up and down for a spin system, or occupied and unoccupied 
for a spinless fermion. The dimension of the Hilbert space
will grow as $2^N$, which makes very hard the study of 
the large $N$ limit, unless some special trick is used. 
The RG method provides a general systematic approach to handle
problems with a large number of degrees of freedom on the 
basic assumption  that only a small 
number of states is needed in order to describe
the long distance physics. How to choose the {\bf{most
representative}} degrees of freedom out of a miriad
of states is the central issue of the RG method.
This can be done in  several manners. 
  We shall introduce below some of them and establish  their comparison.

The DMRG method was originally formulated as
a real space RG method although it admits also
a momentum space formulation \cite{Xiang}.
 We next introduce 
the basic concepts common to any real space RG method
and later on we confine ourselves to the DMRG.

The real space RG consist essentially of 3 steps: i) blocking,
ii) truncation and iii) iteration. First one divides 
the system into blocks, then one finds an effective
description of these blocks in terms of  intra-block and
inter-block interactions and finally one iterates the
algorithm. One can distingish between  3 types of blockings
Kadanof-blocking, Wilsonian-blocking and DMRG-blocking.

\subsection*{Kadanof Blocking}

We shall first consider the case of a linear chain
with $N$ sites. In the first step of the Kadanof blocking
one divides the system into blocks of 2 sites, thus for a
$N=16$ chain one gets,

\begin{equation}
(\bullet \; \bullet) \; ( \bullet \; \bullet) \;
(\bullet \;  \bullet) \; ( \bullet \; \bullet)\;
(\bullet \; \bullet) \; ( \bullet \; \bullet) \;
(\bullet \;  \bullet) \; ( \bullet \; \bullet)
\label{1}
\end{equation}

If every site describes two states then
 the block $(\bullet \; \bullet)$
describes 4 states. Eq.(\ref{1}) is nothing but
a change from a 1-site basis 
to a 2-sites basis and hence $\bullet \; \bullet$ is
entirely equivalent to $(\bullet \; \bullet)$. 
The goal of eq.(1) is to  prepare the road to perform 
the first RG truncation. Indeed 
out of the 4 states we may already want
to keep a smaller number, say 3, 2 or even 1. 
We can represent symbolically this operation as follows,

\begin{equation}
(\bullet \; \bullet) \rightarrow (\bullet \; \bullet)' \;\,\, 
(\rm{truncation})
\label{2}
\end{equation}

From a formal point of view the  blocking operation 
is captured by putting parenthesis $( \dots )$ around the
sites subjected to the blocking while the truncation operation
is represented by $'$ acting on the corresponding block.
Combining eqs. (\ref{1}) and (\ref{2}) the chain with $N$ sites
become after
the first blocking and truncation,

\begin{equation}
(\bullet \; \bullet)' \; ( \bullet \; \bullet)' \;
(\bullet \;  \bullet)' \; ( \bullet \; \bullet)'\;
(\bullet \; \bullet)' \; ( \bullet \; \bullet)' \;
(\bullet \;  \bullet)' \; ( \bullet \; \bullet)'
\label{3}
\end{equation}

If only one state is kept in (\ref{3})  then 
the RG process ends up since there is already a
single state to represent the ground state of the system.
A dimerized spin chain is a typical example of this
type of states, where $(\bullet \; \bullet)'$ is
given by the singlet formed by two spin 1/2 of the chain.

In general however  one keeps more than one state per block 
 $(\bullet \; \bullet)'$ and so one can continue
the RG method choosing  $(\bullet \; \bullet)'$ as 
a the new  site $\bullet'$, i.e.

\begin{equation}
(\bullet \; \bullet)' \rightarrow \bullet' 
\label{4}
\end{equation}

\noindent

The renormalized  chain has therefore  $N/2$ effective
sites $\bullet'$, which can again be blocked as in (\ref{1}),

\begin{equation}
((\bullet \; \bullet)' \; ( \bullet \; \bullet)')' \;
((\bullet \;  \bullet)' \; ( \bullet \; \bullet)')'\;
((\bullet \; \bullet)' \; ( \bullet \; \bullet)')' \;
((\bullet \;  \bullet)' \; ( \bullet \; \bullet)')'
\label{5}
\end{equation}

Performing 
two more blockings and truncation operations we finally get

\begin{equation}
((((\bullet \; \bullet)' \; ( \bullet \; \bullet)')' \;
((\bullet \;  \bullet)' \; ( \bullet \; \bullet)')')'\;
(((\bullet \; \bullet)' \; ( \bullet \; \bullet)')' \;
((\bullet \;  \bullet)' \; ( \bullet \; \bullet)')')')'
\label{6}
\end{equation}

Eq.(\ref{6}) is the final step of the RG method since the
whole chain with $N$ sites has been reduced to a single
effective site whose dynamics can a priori be easily found
by solving the  final renormalized  effective  Hamiltonian.

\subsection*{Wilsonian Blocking}

In his solution of the Kondo impurity problem Wilson \cite{wilson}
introduced a numerical RG method where a single block
is grown by adding momentum shells following the so
called onion scheme.
The real space version of this method is summarized
in the following eq.

\begin{equation}
(((((((((((((((\bullet \; \bullet)' \;  \bullet)' \; \bullet)' \;
\bullet)' \;  \bullet)' \;  \bullet)' \; \bullet)'\;
\bullet)' \; \bullet)' \;  \bullet)' \; \bullet)' \;
\bullet)' \;  \bullet)' \;  \bullet)' \; \bullet)'
\label{7}
\end{equation}

\noindent where we have used the same notations as for the
Kadanof blocking.  While the Kadanof blocking 
follows the pattern $B_{\ell} B_{\ell} \rightarrow
B'_{2 \ell}$, the Wilsonian scheme follows the pattern
$B_\ell \; \bullet \rightarrow B'_{\ell+1}$, where
$B_\ell$ denotes a block with $\ell$ sites.
It seems that the two schemes are completely  unrelated. 
Notice however  that the number of left and right parenthesis in 
eqs. (\ref{6}) and (\ref{7}) is the same, namely $N-1=15$. 
What is different is the order of the brackets. 
The condition for the Kadanof and Wilsonian blockings to be equivalent
can be formulated as follows,

\begin{equation}
((B_1 \; B_2)'\; B_3)' = ( B_1 \;( B_2 \; B_3)')'
\label{8}
\end{equation}

\noindent
where $B_i (i=1,2,3)$ denote 
generic blocks containing one or more sites.
In particular for $N=4$  one can prove using (\ref{8})

\begin{equation}
((\bullet \; \bullet)' \;  (\bullet \; \bullet)')' =
(((\bullet \;  \bullet)' \;  \bullet)' \; \bullet)'
\label{9}
\end{equation}

Eq.(\ref{8}) is reminiscent of the associativity of the
tensor product of representations in group theory and more
precisely in quantum group theory ( see below). This
equation certainly holds if there is no truncation
of degrees of freedom, i.e. $(B_1 \; B_2)' = (B_1 \; B_2)$
in which case it amounts to the equivalence between different
basis. In group theory the relation
between different basis  is given by $6-j$ symbols.

Quantum groups are $q$-deformations of classical
groups ( $q=1$ in this notation) where some
of the commutator and addition rules are deformed 
(for a review see \cite{q-book}). 
The representation theory of quantum groups,
when the deformation parameter $q$ is generic
is analogue to that of classical groups.
However when $q$ is a root of unit
things change completely. First of all
there are a finite number of regular
irreps and 
the tensor product of them 
is also truncated while keeping the associativity 
condition (\ref{8}). 

The existence of an associative truncated tensor
is a common feature of the DMRG, quantum groups and Conformal
Field Theory (CFT) (more on this point below).

\subsection*{DMRG Blocking}

There are two DMRG algorithms to study  open 
chains. The infinite system algorithm uses
the superblock $B_{\ell} \; \bullet \; \bullet B^R_\ell$
to grow the chain from both sides according
to the Wilsonian scheme \cite{W1}. The 
block $B_\ell \bullet$ is then truncated to
a new block $B'_{\ell +1}$. In this manner 
the size of the system grows indefinitely
until one reaches a fixed point beyond
which the numerical results reproduce themselves.
This method is very good in computing bulk
properties of the system like the ground state
energy density. In many cases however it is more convenient
to study finite size systems whose large
distance properties can later
on be obtained trough finite size scaling techniques.
This is notably the case of gapless systems. 

The DMRG algorithm used in these cases is called
the finite system method and it is extremely accurate.
The first steps of this method uses the infinite system
algorithm to grow both sides of the chain independently
until the left and right blocks are  a half  the size of the chain. 
The chain with $N$ (even) sites 
is then obtained by joining a left block  $B_{N/2}$  
and a right block $B_{N/2}^R$ as follows (the superindex 
$R$ in $ B_{N/2}^R $ indicates that it can be obtained from the reflection 
of a righ block),

\begin{equation}
((((((((\bullet \; \bullet)' \;  \bullet)' \; \bullet)' \;
\bullet)' \;  \bullet)' \;  \bullet)' \; \bullet)'
(\bullet  \; (\bullet \;  (\bullet \; (\bullet \;
(\bullet \;  (\bullet \;  (\bullet \; \bullet)')')')')')')')'
\label{10}
\end{equation}

The superblock (\ref{10}) is used in the DMRG to 
enlarge  the left block from $B_{N/2}$ to  $B_{N/2 +1}$
while the right block is reduced from $B^R_{N/2}$ to
$B^R_{N/2-1}$, in  which case we get

\begin{equation}
(((((((((\bullet \; \bullet)' \;  \bullet)' \; \bullet)' \;
\bullet)' \;  \bullet)' \;  \bullet)' \; \bullet)'
\bullet)'  \; (\bullet \;  (\bullet \; (\bullet \;
(\bullet \;  (\bullet \;  (\bullet \; \bullet)')')')')')')'
\label{11}
\end{equation}

If the associativity eq.(\ref{8}) would hold then 
the blockings (\ref{10}) and (\ref{11}) would give
the same result for the GS of the whole chain 
 but of course this is not  the case. The next step
is to again enlarge the left  block at the expenses
of the right one until one reaches the right hand side.
There one reverses the trend and grows the right blocks
at the expenses of the left ones. After several sweeps
of this back-and-forth algorithm 
the GS energy and  GS wave function converge to a fixed
values  which are independent of the size
of left and right blocks.

In this moment the splitting of the chain into left and right
blocks is independent of their size  so that 
the associativity constraint ( \ref{8})
is effectively fullfilled.

The analysis performed so far is rather formal but
helps to  abstract the blocking 
procedure which is common to all the real space RG methods.  
As a by-product we have shown that 
the blocking and the iteration procedures  have to 
be considered as combined strategies to achieve
the same goal which is to reduce the whole 
system to a single effective site. From a formal
point of view blocking is like tensoring representations.
In this sense the RG steps can be seen as 
``putting parenthesis''  around the blocks. An exact RG method
would be the one for which the final result would be independent
on the way the parenthesis are put on. These lead us to the associativity
constraint (\ref{8}), whose fullfillement is the actual goal
of any exact RG method.

\section*{The Standard RG Algorithm}

In all the real space RG methods
there is  an algorithm to truncate the collection of two blocks 
$B_1 B_2$ down to
a new effective block  $ B'_{12}$ where 
$B_1$ or $B_2$ may stand also for a single site.

The standard RG algorithm consist of the following steps
1) diagonalization of the Hamiltonian $H_{B_1 B_2}$
for the combined
block $B_1 B_2$, 2) truncation to the lowest energy
states of $H_{B_1 B_2}$ and 3) change of basis to
the new states kept and renormalization of the old
Hamiltonian.

This method leads to a lot of problems whose  
origin was first pointed out in reference \cite{pinbox}
following a suggestion by Wilson. Studying 
the very simple problem of a particle in a box
the authors of reference \cite{pinbox} interpreted
the bad performance of the standard RG 
as been due to an incorrect treatment of the boundary
conditions applied on a block by its neighbours. 
In other words the truncation $B_1 B_2 \rightarrow
(B_1 B_2)'$ has to take into account the presence
of say a third block $B_3$ in contact with the former
ones. The key idea is to consider a superblock 
$B_1 B_2 B_3$ where the effect of $B_3$ into 
the other two blocks can be properly 
considered. 
An alternative RG-solution to the particle-in-a-box problem, 
which also takes into account the effect of boundary conditions
has been given in \cite{role}.

\section*{The DMRG Algorithm}

Let us choose a superblock made out of three
blocks $B_1 B_2 B_3$. The middle block
is  taken to be a single site or two sites
$B_2 = \bullet$ or $\bullet \; \bullet$. Then  one
constructs the Hamiltonian $H_{B_1 B_2 B_3}$
describing the dynamics of the superblock and finds out 
a given state called the target state, which is usually
the ground state of the superblock
which can be written as

\begin{equation}
|\psi \rangle = \sum_{i_1, i_2, i_3} \psi_{i_1 i_2 i_3}
\; |i_1, i_2, i_3 \rangle  
\label{12}
\end{equation}

\noindent where $i_1,\dots$ run from 1 up to $m_1, \dots$. 
The superblock can be regarded 
either as $((B_1 B_2) B_3)$ or as $(B_1 (B_2 B_3))$. 
Correspondingly the target wave function 
can be written in two different manners,

\begin{eqnarray}
& \psi_{i_1 i_2 i_3} = \sum_{\alpha} \,
U_{i_1 i_2, \alpha} \,D^{(12)3}_\alpha \, V_{\alpha,i_3} & \label{13} \\
& \psi_{i_1 i_2 i_3} = \sum_{\beta} \,
U_{i_1, \beta} \,D^{1(23)}_\beta \, V_{\beta, i_2 i_3} & \nonumber 
\end{eqnarray}

\noindent where $U$ and $V$ are matrices which ``diagonalize'' the wavefunction
and satisfy the orthogonality conditions,

\begin{eqnarray}
& \sum_{i_1 i_2} \; U^*_{i_1 i_2, \alpha} \;  U_{i_1 i_2, \alpha'} 
= \delta_{\alpha, \alpha'} & \label{14} \\
& \sum_{i_3} \; V^*_{\alpha, i_3} \; V_{\alpha', i_3} 
= \delta_{\alpha, \alpha'} & \label{15} 
\end{eqnarray}

We have used in 
(\ref{13}) the singular value 
decomposition (SVD) of a matrix \cite{W1}. 
$D^{(12)3}$ and $D^{1(23)}$ are the singular values of 
$\psi$ regarded as a $(m_1 m_2) \times m_3$ or as a  $m_1 (m_2 \times m_3)$
matrix. Eqs.(\ref{13}) are the clue of the DMRG method.
Let us imagine for a moment that $D_\alpha^{(12)3}$ and 
$D^{1(23)}_\beta$ are zero for certain values of $\alpha$ and
$\beta$. In this case it is clear that we can truncate 
the states of $B_1 B_2$ ( resp. $B_2 B_3$) down to a smaller set of states
$\alpha$ (resp. $\beta$)
for which $D^{(12)3}_\alpha$ ( resp. $D^{1(23)}_\beta$) is non zero
without loosing any information in order to reconstruct
the target state $\psi$. Rather than performing the SVD 
of $\psi$ it is more convenient to define the density
matrices for the subsystems $(12)$ and $(23)$ inside
the whole system $(123)$ \cite{W1},

\begin{eqnarray}
& \rho_{i_1 i_2, i_1' i_2'}^{(12)} = 
\sum_{i_3} \; \psi^*_{i_1 i_2 i_3} \; \psi_{i_1' i_2' i_3} & \label{16} \\ 
& \rho_{i_2 i_3, i_2' i_3'}^{(23)} = 
\sum_{i_1} \; \psi^*_{i_1 i_2 i_3} \; \psi_{i_1 i_2' i_3'} & \nonumber 
\end{eqnarray}

Now using (\ref{13}), (\ref{15}) we get,

\begin{eqnarray} 
& \rho_{i_1 i_2, i_1' i_2'}^{(12)} = 
\sum_{\alpha} \; U^*_{i_1 i_2, \alpha} \; \left( D^{(12)3}_\alpha \right)^2
\; U_{i_1' i_2', \alpha} &  \label{17} \\
& \rho^{(23)}_{i_2 i_3, i'_2 i'_3} = \sum_{\beta}
\; V^*_{\beta,i_2 i_3} \; \left( D^{1(23)}_{\beta} \right)^2
\; V_{\beta, i'_2 i'_3} & \nonumber  
\end{eqnarray}

Eqs.(\ref{17})  means that $w_\alpha^{(12)} = 
(D^{(12)3}_\alpha)^2$ are the eigenvalues
of the  density matrix $\rho^{(12)}$ while $U$ is the unitary matrix
which diagonalizes $\rho^{(12)}$ (similar properties hold
for the density matrix $\rho^{(23)}$.)  
Let us call $m$ the number of states kept per block in a DMRG computation.
This number typically varies between  10 and  1000 depending on the computer
resources.

The DMRG algorithm consists in choosing the $m$ most
probable states $\alpha$, i.e. the states with higher value of
$w_\alpha$ ( let us sort them as 
 $(w_1 \geq w_2 \geq w_3 \geq \dots \geq w_{m_1 m_2}$)
This guarantees 
the best posible representation of
the target state $\psi$ for every given value of $m$. 
Moreover the sum 
$P(m)= \sum_{\alpha=1 }^{m} w_\alpha$
of the probabilities of the $m$ states kept give a reasonable measure
of the  truncation error (recall that tr$ \rho^{(12)}= \sum_\alpha
w_\alpha =1$ and hence $P(m) \leq 1$).

In many of the 1d models studied with the DMRG it turns out that
the probability $w_\alpha$ is concentrated in a few states and
that it decays exponentially fast. This implies that with small
values of $m$ one can achieve a great accuracy in representing 
the target state. This is certainly the case for systems with
a finite correlation length\cite{OR,Jorge,JM}.
For systems with an infinite correlation length 
one has to study finite systems and adjust the number
of states kept $m$ to the correlation length due
the finite size\cite{ABO}.

\section*{The DMRG versus  Quantum Group Theory and CFT}

There are certain formal analogies between the DMRG and the  
theory of quantum groups and Conformal Field
Theories (CFT) which we shall review below. 
First of all
the DMRG truncation of states in $(B_1 B_2)'$ has strong
similarities with the truncated tensor product of irreps
of a $q$-Group where $q$ is a root of unit \cite{duality}. Let us choose
for example the quantum group $SU(2)_q$ which is a 
$q$-deformation of the rotation group $SU(2)$.
For generic values of $q$ the representation theory
of $SU(2)_q$ is similar to that of $SU(2)$, i.e.
every irrep corresponds to an integer or half integer 
spin $j=0, 1/2, \dots$ and the tensor product 
of irreps satisfies  the standard Clebsch-Gordan 
decomposition. However, if $q$ is a root of unit, 
$q= e^{2 \pi i/(k+2)} $ then there is only a  finite
number of regular irreps corresponding to the spins
$j=0, 1/2, \dots, k/2$. The  tensor product
 of these irreps is 
a truncated version of the classical CG decomposition,

\begin{equation}
(V_{j_1} \otimes V_{j_2} )'= 
\oplus_{j=|j_1-j_2|}^{{\rm min}(j_1+j_2, k-j_1-j_2)}
\; V_j   
\label{18}
\end{equation}

\noindent $V_j$ denotes the vector space of dimension
$2j+1$ associated to the irrep with spin $j$. 
It is interesting to observe that the truncated
tensor product (\ref{18}) satisfies the associativity
condition (\ref{8}), namely \cite{duality},

\begin{equation}
((V_1 \otimes V_2)' \otimes V_3)' = ( V_1 \otimes( V_2 \otimes V_3)')'
\label{19}
\end{equation}

This eq. is a consequence of the co-associativity of the
comultiplication of the quantum group $SU(2)_q$. In more 
physical terms,  eq.(\ref{19}) follows from the non trivial 
addition rule of angular momenta in $SU(2)_q$.
The regular irreps have positive $q$-dimension which is defined
as \cite{duality},

\begin{equation}
d_j \equiv [2j+1]_q \equiv \frac{q^{(2j+1)/2} - 
q^{-(2j+1)/2}}{q^{1/2}- q^{-1/2}}
\label{20}
\end{equation}

The $q-$dimension of an irrep plays a role similar to
the eigenvalues $w_\alpha$ of the density matrix 
in the sense that irreps with zero $q$-dimension
are thrown away in the tensor product just like
in the DMRG truncation. 
Based on this analogy we conjecture
that a $q$-group invariant Hamiltonian, like
the XXZ open chain with $q$ a root of unit, 
when studied with
DMRG methods will yield a density matrix with vanishing 
eigenvalues corresponding to non regular irreps. 
The DMRG truncation of these states have to
agree with the $q$-group truncation of the
non regular states \cite{q-group}.

Quantum groups with $q$ a root of unit are 
intimately related to rational CFT's (RCFT).
Indeed in a RCFT 
there is a finite number of primary fields $\phi_a
( a= 1, \dots, M)$, 
which are in one-to-one correspondence
with the regular irreps of the associated $q$-group \cite{duality}.  
Hence from the previous relation between the DMRG
and $q$-groups we may expect a relationship between
RCFT's and the DMRG. More generally 
in a CFT there  are 
null states in the Verma modules of the primary
fields, whose norm is  zero. As shown by
Belavin, Polyakov and Zamolodchikov (BPZ)  the 
decoupling of null vectors 
leads to a set of partial differential equations
for the conformal blocks of the theory, in terms
of which one can construct all the correlators
of the theory \cite{BPZ}. It is tempting to suggest that
the BPZ  decoupling of null vectors is 
the field theoretical  version  of the DMRG truncation.
On the other hand 
the analogue of the tensor product decomposition
is given by the fusion rules of the primary fields,

\begin{equation}
\phi_a \otimes \phi_b = N_{a, b}^c \; \phi_c 
\label{fusion}
\end{equation}

\noindent where $N_{a,b}^c$ is an integer which counts
how many times  the primary field $\phi_c$ appears into
the Operator Product Expansion (OPE) of $\phi_a$ and $\phi_b$. 
The associativity of the OPE, i.e.

\begin{equation}
((\phi_a \otimes \phi_b) \otimes \phi_c) = 
(\phi_a \otimes (\phi_b \otimes \phi_c)) 
\label{associa}
\end{equation}

\noindent 
implies a non trivial eq. for the fusion coefficients $N_{a,b}^c$
namely,

\begin{equation}
\sum_d \; N_{a,b}^d \; N_{d,c}^f = \; 
\sum_d \; N_{a,d}^f \; N_{b,c}^d
\label{a-fusion}
\end{equation}

An example of RCFT is given by the $SU(2)_k$ WZW model
with level $k$ \cite{su2}. The primary fields $\phi_j$  are labelled by the
spin $j=0,1/2,\dots, k/2$, while the fusion rules are given by the 
eq.(\ref{18}) with the translation $V_j \rightarrow \phi_j$.
Indeed, as shown in reference \cite{duality}, there is a one-to-one
correspondence between the $SU(2)_k$ WZW model and the quantum
group $SU(2)_q$.

Another aspect of the relation between CFT, Integrable Systems  and the DMRG
concerns the explanation of 
the exponential decay of the 
eigenvalues of the density matrix.
An approach to study this connection 
is through the 
relation between the DMRG density matrix and the
Corner Transfer Matrix (CTM) of Baxter first pointed out by
Nishino \cite{nishino1} in his application 
of the DMRG to classical statistical mechanical
models in 2D. As shown by Baxter \cite{Baxter}
 in an integrable system the eigenvalues of the CTM 
have a very simple structure, i.e. they go as $a^n $
with $n$ an integer. One can recognize here the
exponential decay of the eigenvalues of the density
matrix \cite{oku},\cite{peschel}.

There are still many aspects to clarify in the relation
between the DMRG and CFT and more generally integrable
systems. This could be a fruitful subject in the near future.

\section*{The DMRG and the  Matrix Product Ansatzs}

The DMRG is a variational method which generates an ansatz  
for the GS state and the excited states. This implies in particular
that the DMRG ground state (GS) energy is an upper bound
of the exact GS energy. The variational ansatz
generated by the DMRG is of the matrix product (MP) type.
This fact was shown by \"{O}stlund and Rommer 
in the thermodynamic limit 
of the DMRG in the case of the spin 1 chain \cite{OR}. These authors
proposed that one could get very good results 
for the GS energy and spin gap by using a MP ansatz
which corresponds to a small value of $m$
in the DMRG. The excitations could also be constructed
as Bloch waves on the MPM state.

To understand why the DMRG gives rise to a MP state
we return to eq.(\ref{13}).  If $|i_1\rangle, |i_2\rangle$
and $|\alpha\rangle$ denote basis of the Hilbert spaces
associated to $B_1, B_2 $ and 
$ (B_1 B_2)'$, then the relation between these
basis is

\begin{equation}
|\alpha\rangle = \sum_{i_1=1}^{m_1}
\sum_{i_2=1}^{m_2}  \; U_{i_1 i_2, \alpha} \; |i_1\rangle \; |i_2\rangle,
\;\;\; (\alpha = 1, \dots, m)
\label{21}
\end{equation}

Since $B_2$ is usually a lattice site $\bullet$ we shall
write eq.(\ref{21}) in the following form,

\begin{equation}
|\alpha\rangle_{N} = \sum_{\beta, s} \; A^N_{\alpha, \beta}[s] \;
|\beta\rangle_{N-1} \; |s_N\rangle
\label{22}
\end{equation}

\noindent where $|s_N\rangle$ denotes the local state associated
to the site located at the $N^{{\rm th}}$ position of the chain,
while $|\alpha\rangle_N$ and $|\beta\rangle_{N-1}$ are the states kept
for the blocks of lengths $N$ and $N-1$ respectively.
$A^N_{\alpha, \beta}[s]$ is a matrix $m \times m$ for each
value of $s$.

Iterating (\ref{22}) until reaching the boundary of the chain
one gets,

\begin{equation}
|\alpha_N, \alpha_0\rangle_N = \left( A^{N}[s_N] \; A^{N-1}[s_{N-1}] \dots
A^{1}[s_1]\right)_{\alpha_N, \alpha_0} \; |s_1\rangle \dots |s_{N-1}\rangle \; |s_N\rangle 
\label{23}
\end{equation}

\noindent where the matrix multiplication of the $A^n[s_n]$ matrices
is implicit. $|\alpha_N, \alpha_0\rangle$ is a collection of states
of an open chain with $N$ sites labelled by the pair 
$(\alpha_N, \alpha_0)$. For a closed chain with periodic 
boundary conditions the ansazt becomes

\begin{equation}
|\psi\rangle_N = {\rm Tr} (A^{N}[s_N] \; A^{N-1}[s_{N-1}] \dots
A^{1}[s_1]) \; |s_1\rangle \dots |s_{N-1}\rangle \; |s_N\rangle 
\label{24}
\end{equation}

A further simplication of (\ref{24}) is to 
assume that all the matrices $A^n[s_n]$
are independent on $n$, i.e.  $A^n[s_n] =  A[s_n]$ ( for all $n$).
This assumption   
can be justified 
in the thermodynamic limit of the DMRG
where it reaches a fixed point \cite{OR}. 
However for finite dimensional systems and specially
for open BC's there will be a non trivial dependence
of $A^n[s_n]$ on $n$. In this sense the DMRG gives 
a non homogenous MP ansatz.

In eq.(\ref{22}) we may want that the states 
$|\alpha\rangle$ form an orthonormal set of states
given that both $|\beta\rangle$ and $|s\rangle$ are orthonormal
sets. This implies the following eq. on $A^N[s]$

\begin{equation}
\sum_{\beta,s} (A^N_{\alpha, \beta}[s])^* \;
A^N_{\alpha', \beta}[s] = \;\; \delta_{\alpha, \alpha'}
\label{25}
\end{equation}

\noindent which is nothing else than the eq. (\ref{14}).  
This eq. expresses the fact that $A^N$ relates orthonormal basis.
But recall that it is not simply a change of basis
because we are truncating states, i.e. $m < m_1 m_2$. 

Given the MP ansatzs (\ref{23}) and (\ref{24}) for open
and closed chains respectively we can use a standard variational 
method to find the amplitudes $A[s]$ which minimize 
the  energy of the ansatz. In references \cite{Jorge,JM}
it was shown that when $N$ is large these minimization 
procedure is similar to the one of the DMRG and that 
in fact there is  a hidden density matrix even though
the algorithm did not try to follow the DMRG method. 
One way to see this is if one define 
the following transfer matrix,

\begin{equation}
T_{\alpha \alpha', \beta \beta'}^N =
\sum_s \; (A^{N}_{\alpha, \beta}[s])^*  \; A_{\alpha', \beta'}^N[s]
\label{26}
\end{equation}

\noindent $T^N$ is a $m^2 \times m^2$ matrix which serves
to relate matrix elements of operators between states
with lengths $N$ and $N-1$, for example

\begin{equation}
_N \langle \alpha| {\cal O} | \alpha'\rangle_N = 
\sum_{\beta, \beta'} \, T_{\alpha \alpha', \beta \beta'}^N 
\; \; _{N-1}\langle \beta| {\cal O}| \beta'\rangle_{N-1} 
\label{27}
\end{equation}

\noindent We are assuming in (\ref{27}) that 
the operator ${\cal O}$ does not act on the $N^{\rm th}$  site. 
The normalization condition (\ref{25}) implies that
$T$ has a right eigenvector with eigenvalue
1 given by $\delta_{\alpha, \alpha'}$, namely

\begin{equation}
\sum_{\beta \beta'} T_{\alpha \alpha', \beta \beta'}^N 
\delta_{\beta, \beta'} =
\delta_{\alpha, \alpha'}
\label{28}
\end{equation}

It then follows that $T$ has a left eigenvector
with eigenvalue 1, i.e.

\begin{equation}
\sum_{\alpha \alpha'} \rho_{\alpha \alpha'}^N
T_{\alpha \alpha', \beta \beta'}^N =
\rho_{\beta, \beta'}^N 
\label{29}
\end{equation}

In references \cite{Jorge,JM}
it was shown that $\rho_{\alpha, \alpha'}^N$ can be identified
with a density matrix and that one is really minimizing the
expectation value of the Hamiltonian 
in the following mixed
state,

\begin{equation}
\rho^N = \sum_{\alpha \alpha'} \rho^{N}_{\alpha \alpha'}
|\alpha \rangle_N \; _N\langle \alpha'|
\label{30}
\end{equation}

From this point of view 
the collection of states $|\alpha\rangle_N$ of the MP method
can be interpreted as the most probable ones that contribute
to the GS wave function of a system with $N + 1 + N =2N +1$ sites.

The important conclusion to be learn from the previous considerations
is that the MP ansatz leads in a natural way to the DMRG algorithm.
This may be interesting regarding further generalizations  of
the DMRG to higher dimensions.

\section*{Matrix Products Ansatzs in 2D}

The DMRG algorithm can be generalized to 
ladders (i.e. collections of a finite number of chains), 
and large 
clusters.
This has been done obtaining  remarkable
results which are difficult to obtain with other 
algorithms \cite{wns}. 
However it has also been shown that the  efficiency 
of the DMRG disminish with 
the width of the system \cite{Liang}. The DMRG algorithm appears
to be essentially one dimensional in the sense that
the RG steps follow a linear pattern no matter
whether the system is 1D or higher dimensional. 
 
Of course any higher dimensional system can be converted into 
a 1D system by allowing  non local interactions. 
However locality seems to be the key of the great
performance of the DMRG in 1D. Hence a truly 
higher dimensional version of the DMRG should 
try to keep locality as a guideline.
That this is in principle possible is suggested
by reference \cite{otsuka} where a DMRG algorithm
is given for a Bethe lattice, whose dimensionality is
actually infinite. 
Also recently, Niggemann et al. \cite{niggemann} have 
constructed a two-dimensional tensor product to 
describe the ground state of a 2D quantum system.
Similarly, Nishino and Okunishi have proposed a
Density Matrix algorithm for 3D classical statistical
mechanics models \cite{nishino3D}.

Another approach is suggested by the equivalence of the
Matrix Product approach and the DMRG for 1d or quasi-1d
systems. We do not know at the moment what is the
formulation of the DMRG in 2D but we do know that 
in 2D there are MP states which were first constructed
by Affleck, Kennedy, Lieb and Tasaki (AKLT) \cite{AKLT}. These
states are valence bond solid states where one
connects local spins through local bonds.

When trying to generalize the 1D MP states 
to 2D we find that there are 
two possible types
of MP states which can be conveniently  named
as vertex-MP and face-MP states, using  standard
Statistical Mechanics terminology \cite{Baxter}.

\subsection*{Vertex-Matrix Product ansatzs}

A vertex model in 2D Statistical Mechanics (SM) 
is a model defined on a square
lattice and such that the lattice variables $i,j, \dots$ live 
on the links while the interaction takes place on the vertices
\cite{Baxter}. The Bolzmann weight thus depends on 4 variables
$W_{i j k l}$ and the whole partition function is obtained
by multiplying the Bolztmann weights of all vertices and
then summing over the values 
of the lattice variables. The 6 vertex and 8 vertex
models are the cannonical examples of these types of models
which have been shown to be integrable.

Motivated by these vertex models  we shall define 
a vertex-MP state 
in terms of a set of amplitudes

\begin{equation}
A_{\alpha, \beta}^{\gamma, \delta}[s] ,\;\; 
(\alpha\,\beta, \dots = 1, \dots, m; \;\; s=1, \dots,m_s)  
\label{2D1}
\end{equation}

\noindent 
where the labels 
$\alpha, \beta, \gamma, \delta$  are associated with the links of the
square lattice while $s$ labels the quantum state, e.g. spin, 
associated to the vertex where the 4-links $\alpha, \beta, \gamma,\delta$ 
meet.  $A_{\alpha, \beta}^{\gamma, \delta}[s]$ is a sort of
Boltzmann weight of a vertex model. The vertex-MP wave function
$\psi(s_1, s_2, \dots, s_N)$ can be obtained by multiplying
all the Boltzmann weights 
$A_{\alpha_i, \beta_i}^{\gamma_i, \delta_i}[s_i]$
and contracting and suming over the links variables according to the same
pattern of a vertex model in Statistical Mechanics \cite{Baxter}. 
Hence the value of the wave function  
$\psi(s_1, s_2, \dots, s_N)$ is given by the partition function
of a vertex model where the Boltzmann weights depend on the value
of the local states $s_i$.
This construction for the square lattice is equivalent to the so
called ``vertex-state representation'' of Niggemann et al. for 
the hexagonal lattice \cite{niggeman}.
  
This construction resembles the one proposed  by Laughlin
concerning the Fractional Quantum Hall effect (FQHE) \cite{hall}. 
More explicitely, Laughlin proposed in \cite{hall} a
variational wave function 
$\psi_m(z_1, \dots, z_N)$ for the ground state of the
$N$ electrons in the lowest Landau level of a 
FQHE with filling factor $\nu=1/m$. The norm $|\psi_m|^2$
of the Laughlin wave function can be interpreted as the
Boltzman weight of a classical one component plasma
constitued by $N$ negative charges of magnitude $m$
in a uniform background of positive charges. The charge
neutrality of the plasma guarantees its stability. 

In our case we also have an associated Statistical Mechanical
model given by the Boltzman weights of a vertex model. 
If we compute the norm of the wave function
$\psi(s_1, \dots, s_N)$, we can  perform the summation
over the ``spin'' indices $s$, in which case the
the norm $\langle\psi| \psi \rangle$ of the vertex-MP state
is  given by  
the partition function of another vertex model whose
Boltzmann weights are defined as,

\begin{equation}
R_{\alpha \alpha', \beta \beta'}^{\gamma \gamma', \delta \delta'}
= \sum_s \; A_{\alpha, \beta}^{\gamma, \delta}[s] \; 
A_{\alpha', \beta'}^{\gamma', \delta'}[s]
\label{2D2}
\end{equation}

This $R$ matrix is the 2D version of the 
$T$ matrix defined in (\ref{26}).
The computation of the norm of 
$|\psi\rangle$ can
be in general a difficult task. However, if the
model defined by the weights (\ref{2D2}) turns
out to be integrable, then we could find the
exact norm in the thermodynamic limit.

The face-MP models can be defined in a similar manner
by a set of variational parameters as in (\ref{2D1}) where
now the variables $\alpha, \dots$ are now associated to the
vertices of the squares while the quantum variable $s$
is associated to the face whose vertices are
$\alpha, \beta, \gamma, \delta$.
This is similar to the face or Interaction Round a Face models (IRF) in 
Statistical Mechanics \cite{Baxter}.

Hence in 2D there are two generic ways to produce MP ansatzs
which are in fact the straightforward generalization of the
1D MP ansatzs. These two generalizations
suggest to use some well
know models as the 6-vertex model to test some of the ideas
presented above.

In summary, we have tried to show in this 
contribution some interesting connections
among seemingly unrelated methods in
condensed matter and field theory. 
Much remains to be done along this direction.

\subsection*{Acknowledgements}

We would like first of all 
to thank J. Dukelsky, T. Nishino, S. R. White, D.J. Scalapino
and J.M. Rom\'an for their collaboration and many discussions on the 
several aspects related to the  work presented in these lectures.

G.S.  would like to thank the organizers of the 
workshop ``Recent Developments in Exact Renormalization Group''
held at Faro, A. Krasnitz, Y. Kubyshin, R. Neves,
R.  Potting and P. S\'{a} 
for their kind invitation to lecture in this meeting and for
their warm hospitality.

We acknowledge support from the DGES under contract PB96-0906.

\end{document}